\newcommand\showchange{TT}
\documentclass[conference]{IEEEtran}
\usepackage{import}
\usepackage{amsthm}
\usepackage{threeparttable}
\usepackage{amsmath}
\usepackage{amssymb}
\newtheorem{definition}{Definition} 

\usepackage{multirow}
\usepackage{enumitem}
\usepackage{xspace}
\usepackage{listings}
\usepackage[misc]{ifsym}
\usepackage{fancyhdr}
\usepackage{pgfplots}
\pgfplotsset{compat=1.12}
\usepackage{subfigure}
\usepackage{filecontents}
\usepackage{caption}
\usepackage{cite}
\usepackage[numbers,sort&compress]{natbib}
\newcommand{\cell}[1]{\begin{tabular}[c]{@{}l@{}}#1\end{tabular}}

\usepackage[ruled,linesnumbered]{algorithm2e} 

\SetAlFnt{\small}
\SetAlCapFnt{\small}
\SetAlCapNameFnt{\small}
\SetAlCapHSkip{0pt}
\IncMargin{-\parindent}
\usepackage{soul}	
\setstcolor{red}

\long\def\com#1{}

\if\showchange

\else

\fi

\newcommand{\blind}[1]{}		

\newcommand{\mysect}[1]{\paragraph{#1}}		
\newcommand{\kw}[1]{{\em#1}}	
\newcommand{\kn}[1]{\texttt{#1}}	

\newcommand\etal{\emph{et al.\ }}
\newcommand\eg{\emph{e.g.,\ }}
\newcommand\ie{\emph{i.e.,\ }}

\theoremstyle{definition}

\newcommand\fullname{{\bf CO}ntext-sensitive and {\bf D}uration-{\bf A}ware 
{\bf R}emapping algorithm~({\sc Codar})\xspace}
\newcommand\name{{\sc Codar} remapper\xspace}
\newcommand\mysys{{\sc Codar}\xspace}
\newcommand\qamfname{{\bf M}ulti-architecture {\bf A}daptive {\bf Q}uantum {\bf A}bstract {\bf M}achine 
 ({\sf maQAM})\xspace}
\newcommand\qam{{\sf maQAM}\xspace}

\usepackage[braket]{qcircuit}
\usepackage{tikz}
\usetikzlibrary{calc}
\usetikzlibrary{backgrounds}
\usetikzlibrary{math}

\newcommand{\Input}[0]{{\bf Input:}}
\newcommand{\Output}[0]{{\bf Output:}}

\newcommand{\Nat}[0]{$\mathbb{N}$\xspace}
\newcommand{\Intmax}[0]{\textsf{INT\_MAX}\xspace}
\newcommand{\Arrow}[2]{{#1 {$\!\rightarrow\!$} #2}}

\newcommand{\NPQubit}[0]{N\xspace}
\newcommand{\NPGate}[0]{M\xspace}
\newcommand{\NLQubit}[0]{n\xspace}
\newcommand{\NLGate}[0]{m\xspace}
\newcommand{\NSeqs}[1]{\{1,2,..., #1\}\xspace}
\newcommand{\QArch}[0]{$\mathbb{A}$\xspace}
\newcommand{\QArchS}[0]{$\mathsf{A_s}$\xspace}
\newcommand{\QArchD}[0]{$\mathsf{A_d}$\xspace}
\newcommand{\LQubit}[0]{$\mathsf{Q_P}$\xspace}          
\newcommand{\PQubit}[0]{$\mathsf{Q_H}$\xspace}          
\newcommand{\PGate}[1]{$\mathsf{G}_{{#1}}$\xspace}      
\newcommand{\lqubit}[1]{{$q_{{#1}}$\xspace}}            
\newcommand{\pqubit}[1]{{$Q_{{#1}}$\xspace}}            
\newcommand{\lqubitvars}[0]{{\lqubit{\NSeqs{\NLQubit}}\xspace}}    
\newcommand{\pqubitvars}[0]{{\pqubit{\NSeqs{\NPQubit}}\xspace}}    
\newcommand{\pgate}[1]{{\textsf{g}$_{\sf {#1}}$\xspace}}    
\newcommand{\lgate}[1]{{$g_{\sf {#1}}$\xspace}}             
\newcommand{\pgatevars}[0]{{\pgate{\NSeqs{\NPGate}}\xspace}}    
\newcommand{\lgatevars}[0]{{\lgate{\NSeqs{\NLGate}}\xspace}}    
\newcommand{\lgates}[1]{{$I_{\sf {#1}}$\xspace}}            
\newcommand{\lgateschd}[2]{{$(g_{{#1}},t_{{#2}})$\xspace}}       
\newcommand{\lgatesschd}[1]{{$\mathcal{E}_{{#1}}$\xspace}}      
\newcommand{\gname}[1]{{\kn{gate(#1)}\xspace}}          
\newcommand{\gqubits}[1]{{\kn{qseq(#1)}\xspace}}        
\newcommand{\QCoupling}[0]{$\mathbb{M}$\xspace}
\newcommand{\QDistance}[0]{$\mathbb{D}$\xspace}
\newcommand{\PQEdge}[0]{$\mathsf{E_H}$\xspace}          
\newcommand{\QCycle}[0]{$\tau_u$}
\newcommand{\GDuration}[1]{{$\tau_{\textsf{#1}}$\xspace}}   
\newcommand{\qprogram}[1]{{$P_{#1}$\xspace}}
\newcommand{\qmap}[1]{{$\pi_{#1}$\xspace}}

\newcommand{\qdepth}[1]{{$h_{{#1}}$\xspace}}

\newcommand{\qlock}{$t_{end}$\xspace}
\newcommand{\Cswap}{$C_{\sf swap}$\xspace}
\newcommand{\Hbasic}{$H_{\sf basic}$\xspace}
\newcommand{\Hfine}{$H_{\sf fine}$\xspace}
\newcommand{\HD}{{\sf HD}\xspace}
\newcommand{\VD}{{\sf VD}\xspace}

\usepackage[normalem]{ulem}
\begin{document}
\title{
\mysys: A Contextual Duration-Aware 
Qubit Mapping for Various NISQ Devices
}
\author{\IEEEauthorblockN{Haowei Deng, Yu Zhang\IEEEauthorrefmark{1} and Quanxi Li}
\IEEEauthorblockA{School of Computer Science and Technology, University of Science and Technology of China, Hefei, China\\
Email: jackdhw@mail.ustc.edu.cn, yuzhang@ustc.edu.cn\IEEEauthorrefmark{1}, crazylqx@mail.ustc.edu.cn
}
}

\small
\maketitle
\begin{abstract}
Quantum computing devices in the NISQ era share common features and challenges like limited connectivity between qubits. Since two-qubit gates are allowed on limited qubit pairs, quantum compilers must transform original quantum programs to fit the hardware constraints. 
Previous works on qubit mapping assume different gates have the same execution duration, which limits them to explore the parallelism from the program. 
To address this drawback, we propose a \qamfname and a \fullname. The \name is aware of gate duration difference and program context, enabling it to extract more parallelism from programs and speed up the quantum programs by 1.23 in simulation on average in different architectures and maintain the fidelity of circuits when running on OriginQ quantum noisy simulator.
\end{abstract}



\section{Introduction}
\label{sec:intro}

Quantum Computing (QC) has attracted huge attention in recent a decade 
due to its ability to exponentially accelerate some important algorithms~\cite{nielsen2010quantum:qcqi}.
Both QC algorithm designers and programmers work at a very high level,
and 
know little about (future) NISQ devices 
that (will) execute quantum programs.
There exists a gap, however, 
between  NISQ devices and the hardware  
requirements (\eg size and reliability) of QC algorithms. 
To bridge the gap, 
QC requires abstraction layer and 
toolchain to translate and optimize quantum programs~\cite{chong2017pl}.
QC compilers typically translate high-level QC code into 
(optimized) circuit-level assembly code in multiple stages.

In order to use NISQ hardware,
quantum circuit programs have to be compiled to the target device,
which includes mapping logical qubits to physical ones of the device.
The mapping step, 
which we focus on in this paper,
faces a tough challenge 
because further physical constraints 
have to be considered.
In fact,
2-qubit gates can only be applied to certain physical qubit pairs.
A common method to solve this problem is to insert
additional SWAP operations in order to ``move''
the logical qubits to positions where they can interact with each other.
This qubit mapping problem has been proved to be a NP-Complete problem~\cite{siraichi2018qubitalloc}.

Previous solutions to this problem 
can be classified into two types.
One of them is to formulate the problem into 
an equivalent mathematical problem and apply a solver
~\cite{venturelli2017temporal,venturelli2018compiling,booth2018comparing,oddi2018greedy,bhattacharjee2017depth,wille2019mapping}
and the other type is to use heuristic search to obtain approximate results
\cite{wille2016look,kole2016heuristic,kole2018new,bhattacharjee2018novel,zulehner2019mapping,li2019tackling}.
The former suffers from a very long runtime 
and can only apply to small-size cases.
The latter is better in runtime, especially when the circuit is large scale. 
 All these algorithms assume that 
different gates have the same execution duration.\par

\begin{table*}[!h]
\centering
\footnotesize
\caption{Parameter information of several quantum computing devices.}  
\label{tab:qc-args}
\begin{tabular}{@{}c|c|c|@{}c|@{}c|@{}c|c|c@{}}
\hline
\multicolumn{2}{c|}{} &  
\multicolumn{2}{c|}{Ion Trap} &
\multicolumn{3}{c|}{Superconducting} & \multirow{2}{*}{Neutral Atom~\cite{sheng2018PRL121.240501}}  
\\
\cline{3-7}
\multicolumn{2}{c|}{}
& Ion Q5~\cite{linke2017QCcomparison}
& Ion Q11~\cite{wright2019benchmarking}
& ~IBM Q5~\cite{linke2017QCcomparison} 
& ~IBM Q16~\cite{murali2019noise}
& IBM Q20~\cite{li2019tackling} 
\\
\hline
\multicolumn{2}{c|}{Available 1-qubit gate}
& \multicolumn{2}{c|}{R$^\theta_\alpha$} &\multicolumn{3}{c|}{X, Y, Z, H, S, T} & $R_{\alpha}^{\theta}$
\\
\multicolumn{2}{c|}{Available 2-qubit gate}
& \multicolumn{2}{c|}{XX} &\multicolumn{3}{c|}{CNOT (CX)} &CNOT (CX)\cite{stockill2017phase}
\\
\hline
\multirow{4}{*}{Fidelity} 
& 1-qubit gate & 99.1(5)\% &99.5\%& 99.7\%  & $\sim$99.8\% & $\sim$99.56\% &99.995\% \cite{sheng2018PRL121.240501}
\\ 
& 2-qubit gate & 97(1)\% & ~97.5\%[95.1\%,98.9\%]& 96.5\% & $\sim$96\% & $\sim$97\% 
& 82\%\cite{maller2015rydberg}
\\ 
& 1-qubit readout & \ket{0}:99.7(1)\%, \ket{1}:99.1(1)\% &99.3\%& $\sim$ 96\% & $\sim$93\% & $\sim$91.2\% & 98.6\% \cite{fuhrmanek2011free}
\\ 
& average readout & 95.7(1)\% &--& $\sim$ 80\% & -- &--  &  $\ge$97.4(3)\% 
\cite{levine2019parallel}\\
\hline

\multirow{2}{*}{Time} & 1-qubit gate & 20$\mu$ s&&130 ns&80 ns& --& 1$\mu$ $\sim$20$\mu$ s
\\
&2-qubit gate & 250$\mu$ s&--& 250-450 ns &170-391 ns& --
& $\sim$10$\mu$ s
\\
\hline

\multicolumn{2}{c|}{Depolarization ($T_1$)}
& $\sim \infty$  &--& $\sim$ 60$\mu$ s  & $\sim$ 70$\mu$ s & 87.29$\mu$ s  & $>$10s
\\
\multicolumn{2}{c|}{Spin dephasing ($T_2$)}
& $\sim$ 0.5s  &--& $\sim$ 60$\mu$ s & $\sim$ 70$\mu$ s& 54.43$\mu$ s & $\sim$ 1s  
\\
\hline
\end{tabular}
\end{table*}

On NISQ hardware,
however,
different gates have different durations (see Table~\ref{tab:qc-args}).
Ignoring the gate duration difference may cause
these algorithms to find the shortest depth but not the shortest execution time.
The real execution time of the circuit is associated with 
the weighted depth, 
in which different gates have different duration weights.
Considering gate duration difference will help the compiler 
make better use of the parallelism of quantum circuits
and generate circuits with shorter execution time.

In this paper,
we focus on solving the \kw{qubit mapping problem} by heuristic search with the consideration of gate duration difference and program context to explore more program's parallelism.
To address the challenges of qubit mapping problem
and adapt to different quantum technologies,
we first give several examples to explain our motivation,
then propose a \qamfname 
for studying the qubit mapping problem.
The \qam is modeled as a 
coupling graph with limited qubit connectivity
and configurable durations of different kinds of quantum gates.
Based on the \qam,
we further propose two mechanisms that enable \fullname to solve the qubit mapping problem with the awareness of gate duration difference and program context.
 \par
The main contributions of this paper are as follows:
\begin{itemize}
    \item We summarize the features of different QC technologies including their available gates, 
    operation fidelity and execution duration,
    and establish the quantum architecture abstraction with configurable parameters -- \qam.    
    \item We propose a SWAP-based heuristic algorithm for qubit remapping, \mysys,
    which considers the gate duration difference and program context and can further speed up the quantum program.
    \item We evaluate \mysys with various benchmarks on several latest hardware models such as Google's 54-qubit Sycamore processor~\cite{arute2019google54}.  
    Experimental results show that \mysys speeds up quantum programs by 1.212$\sim$1.258 at the average in comparison with the best-known algorithm, and maintains the fidelity of circuits when running on OriginQ quantum noisy simulator~\cite{qpanda}. 
\end{itemize}\par
\section{Problem Analysis}
\label{sec:problem}

\subsection{Recent Work on Qubit Mapping}
\label{sec:problem:related}
There are a lot of research on the qubit mapping problem.
Here we focus on analyzing some valuable solutions in recent two years~\cite{siraichi2018qubitalloc,zulehner2019mapping,li2019tackling,wille2019mapping,murali2019noise,ash2019qure,tannu2019not}.
All of them 
are proposed for 
some IBM QX architectures,
and none of them consider the gate duration difference.


\paragraph{Solutions only considering qubit coupling}
\cite{siraichi2018qubitalloc,wille2019mapping} provide solutions for 5-qubit IBM QX architectures with directed coupling.
Siraichi \etal~\cite{siraichi2018qubitalloc} propose an optimal algorithm 
based on dynamic programming, 
which only fits for small circuits;
then they propose a heuristic one which is fast but oversimplified with results worse than IBM's solution.
Wille \etal~\cite{wille2019mapping} present a solution with a minimal number of additional \kn{SWAP} and \kn{H} operations,
in which qubit mapping problem is formulated as a symbolic optimization problem with high complexity. 
They utilize powerful reasoning engines to solve the computationally task.

\cite{zulehner2019mapping,li2019tackling} use heuristic search to provide good solutions 
in acceptable time for large scale circuits.
Zulehner \etal\cite{zulehner2019mapping} divide the two-qubit gates into independent layers, 
then use $A^*$ search plus heuristic cost function to determine compliant mappings for each layer.
Li \etal\cite{li2019tackling} propose a SWAP-based bidirectional heuristic search algorithm -- SABRE, 
which can produce comparable results with exponential speedup
against previous solutions such as~\cite{zulehner2019mapping}. 


\paragraph{Solutions further considering error rates}
\cite{ash2019qure,murali2019noise,tannu2019not} provide another type of perspective for solving the qubit mapping problem. 
They consider the variation in the error rates of different qubits and connections to generate directly executable circuits that improve reliability rather than minimize circuit depth and number of gates. 
Based on the error rate data from real 
IBM Q16 and Q20 respectively,
\cite{murali2019noise,tannu2019not} 
use a SMT solver to schedule gate operations to qubits with lower error probabilities. 
Ash-Saki \etal propose two approaches, Sub-graph Search and Greedy approach, to 
optimize gate-errors~\cite{ash2019qure}.
Circuits generated by them may suffer from long execution time due to no consideration of the minimal circuit depth.

\paragraph{What we consider in the qubit mapping}
We want to produce solutions for the qubit mapping problem with speedup against previous works and maintain the fidelity meanwhile. 
Besides the coupling map, 
what we further concern includes the program context and the gate duration difference,
which affect the design of qubit mapping.
Considering these factors will help to find remapping solution with approximate optimal execution close to reality.

\subsection{Motivating Examples}
\label{sec:problem:ex}
\begin{figure}[h]
\begin{minipage}[b]{.12\textwidth}
\centering
{
\includegraphics[width=1\textwidth]{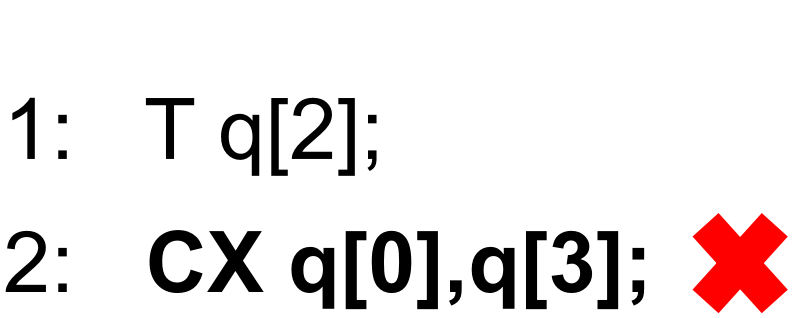}
\includegraphics[width=0.65\textwidth]{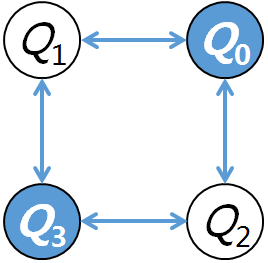}
\\~~
{(a)}
}
\end{minipage}
~
\begin{minipage}[b]{.35\textwidth}
\centering
\includegraphics[width=\textwidth]{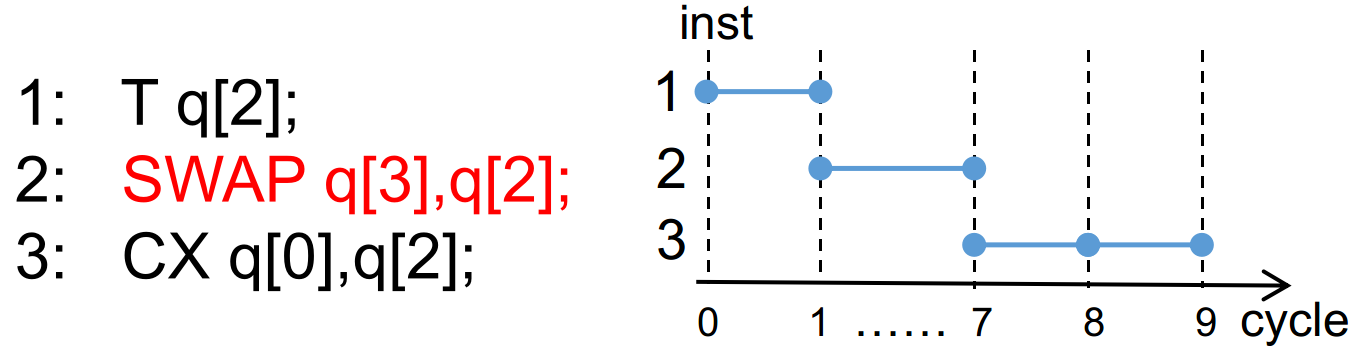}
\\~~
{(c)}
\end{minipage}
\\
\begin{minipage}[b]{.12\textwidth}
\centering
\scriptsize
\begin{tabular}{c|c}
\hline
    Gate  & Duration
    \\
    \hline
    T & 1 cycle \\
    \hline
    CX & 2 cycle \\
    \hline
    SWAP & 6 cycle \\
    \hline
\end{tabular}
\normalsize
\\~~\\~~
{(b)}
\end{minipage}
~
\begin{minipage}[b]{0.35\textwidth}
\centering
\includegraphics[width=\textwidth]{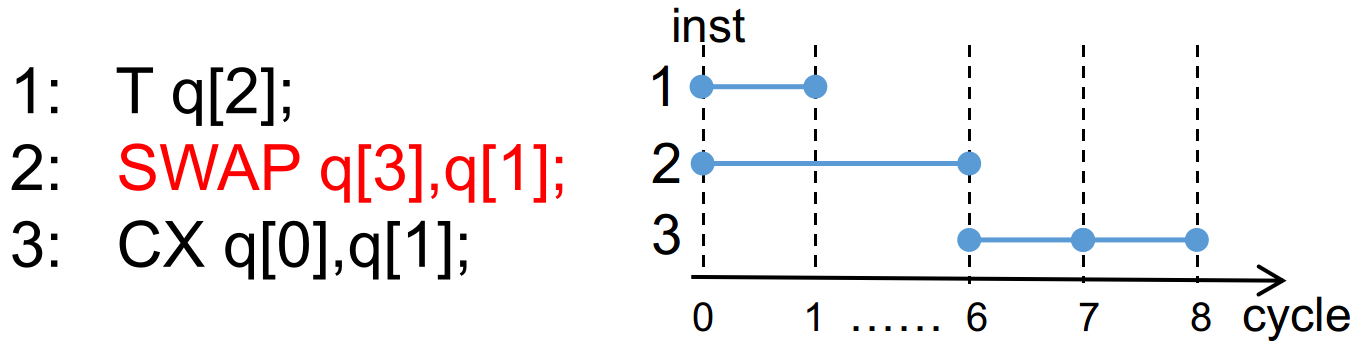}
\\
{(d)}
\end{minipage}
\caption{An example reflecting the impact of {\bf program context} on SWAP-based transformations: 
\kn{SWAP q[3], q[1]} is selected in (d) to avoid using \kn{q[2]} operated by the previous \kn{T} gate, accordingly increasing parallelism.}
\label{fig:ex1:qft4}
\end{figure}

We use several examples written in   
OpenQASM~\cite{cross2017OpenQASM}   
to explain our motivation for considering program context
and gate duration difference in the qubit remapping process.
The two examples base on the coupling map of four physical qubits $Q_0\sim Q_3$ and 
the assumed gate durations 
defined in Fig.~\ref{fig:ex1:qft4} (a) and (b).
We directly map the logical qubits \kn{q[0]}$\sim$\kn{q[3]} initially 
to physical qubits $Q_0\sim Q_3$ for easier explanation.

\begin{figure}[h]
\begin{minipage}[b]{.15\textwidth}
\centering
\includegraphics[width=0.5\textwidth]{figs/grid4.png}
\\
{(a)}
\end{minipage}
\begin{minipage}[b]{.35\textwidth}
\centering
\includegraphics[width=0.9\textwidth]{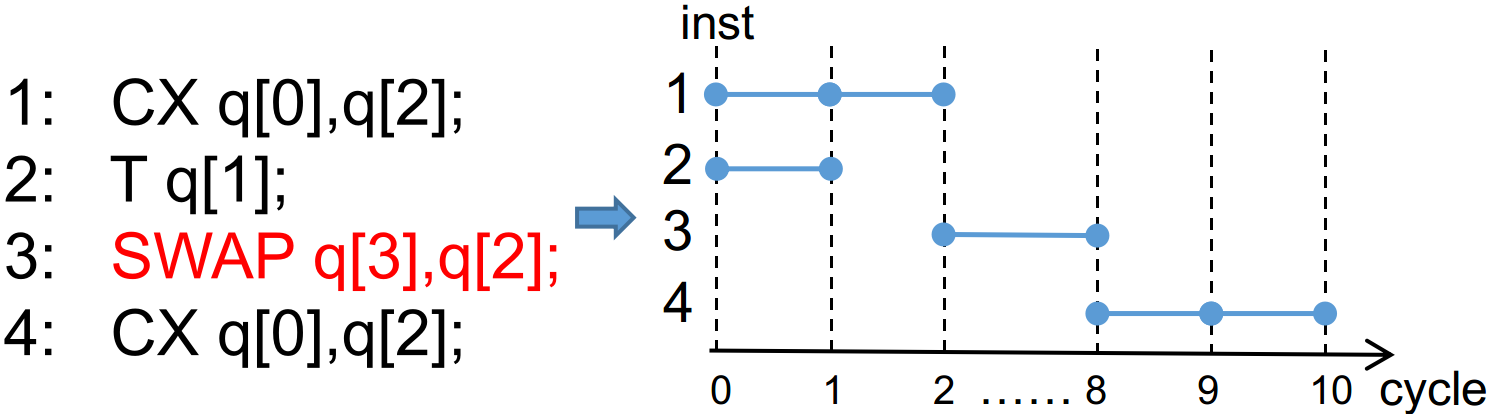}
\\
{(c)}
\end{minipage}
\\~~\\
\begin{minipage}[b]{0.15\textwidth}
\centering
\includegraphics[width=0.9\textwidth]{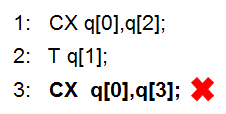}
{(b)}
\end{minipage}
~
\begin{minipage}[b]{0.35\textwidth}
\centering
\includegraphics[width=0.9\textwidth]{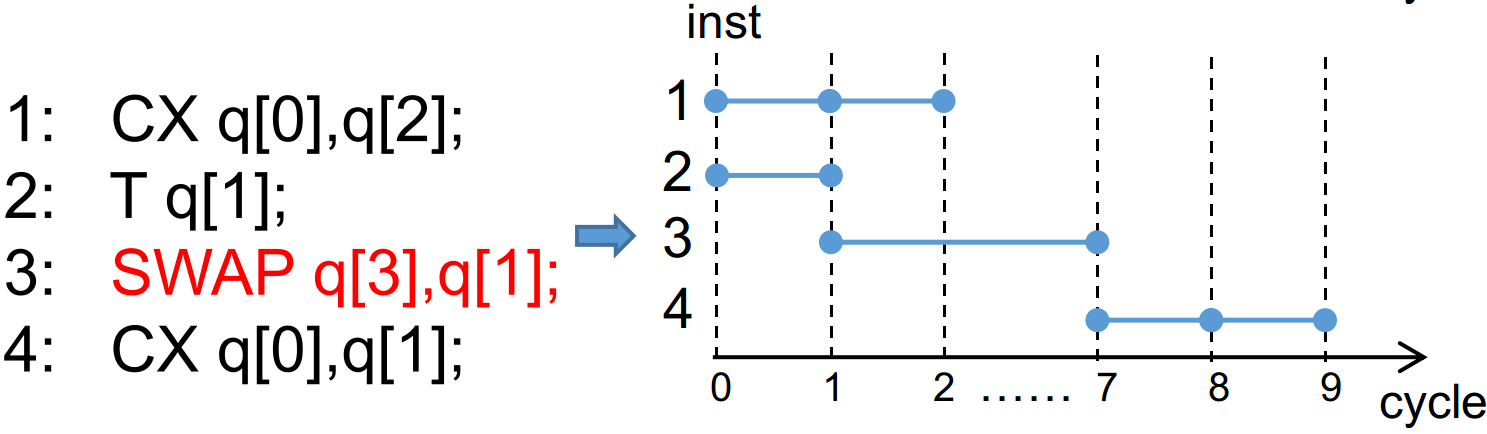}
\\{(d)}
\end{minipage}
\caption{A 4-qubit QFT example reflecting the impact of {\bf gate duration difference}: 
``\kn{SWAP q[3],q[1]}'' 
is the best candidate
since it can start immediately after
``\kn{T q[1]}'' while ``\kn{CX q[0],q[2]}'' has not finished yet, increasing the parallelism of the circuit.}
\label{fig:exdu:qft4}
\end{figure}

\mysect{Impact of program context}
\label{sec:problem:ex:context}


Consider the OpenQASM code fragment 
shown in Fig.\ref{fig:ex1:qft4} (b).
Since qubits $Q_0$ and $Q_3$ are non-adjacent,
the instruction ``\kn{CX q[0],q[3]}'' at line 2 cannot be applied. 
To solve the problem,
\kn{SWAP} operation is required before performing the \kn{CX} operation.
In this case,
there are four candidate \kn{SWAP} pairs,
\ie ($Q_0$, $Q_1$), ($Q_0$, $Q_2$), ($Q_3$, $Q_1$) and ($Q_3$, $Q_2$). 
If the program context, \ie the predecessor instruction 
``\kn{T q[2];}'', is not considered, 
there are no differences among the four candidates when selecting.
However, \kn{SWAP} operation on pair ($Q_3$, $Q_2$) or ($Q_0$, $Q_2$) conflicts with the context instruction ``\kn{T q[2];}'' due to operating the same $Q_2$,
and has to be executed serially after \kn{T} operation as shown in Fig.\ref{fig:ex1:qft4} (c).
\kn{SWAP} on pair ($Q_3$, $Q_1$) or ($Q_3$, $Q_1$) does not conflict with ``\kn{T q[2];}'' 
and can be executed in parallel as shown in Fig.\ref{fig:ex1:qft4} (d).
With the awareness of the context information, 
\kn{SWAP} operations that improve parallelism can be sifted out.

\mysect{Impact of gate duration difference}
\label{sec:problem:ex:duration}
We use a 4-qubit QFT (Quantum Fourier Transform) circuit to explain the limitation of ignoring the duration of quantum gates.
Fig.~\ref{fig:exdu:qft4} (b) lists the fragment of a 4-qubit QFT OpenQASM program,
which is generated by ScaffCC compiler~\cite{abhari2014scaffcc}.
Similar to the first example, 
\kn{SWAP} operation is required before performing the \kn{CX} operation, and there are also the same four candidate \kn{SWAP} pairs. 
Instructions ``\kn{T q[2]}'' and ``\kn{CX q[0],q[2]}'' can be executed in parallel 
and we assume both of them start at cycle 0. 
If the difference of gate durations is ignored, 
the two gates ``\kn{T q[2]}'' and ``\kn{CX q[0],q[2]}'' are assumed to finish at the same time $t$ 
and the four candidate \kn{SWAP} operations have to start after $t$. 
However, if the duration of \kn{CX} is twice as much as that of \kn{T},
we find that ``\kn{T q[2]}'' will finish at cycle 1 while ``\kn{CX q[0],q[2]}'' at cycle 2. 
As a result, \kn{SWAP} between \kn{q[3]} and \kn{q[1]} can start at cycle 1 as shown in Fig.\ref{fig:exdu:qft4} (d),
while other three candidate SWAP operations 
have to start at cycle 2
since one of operands $Q_0$ or $Q_2$ is occupied
as shown in Fig.\ref{fig:exdu:qft4} (c). 
Fig.~\ref{fig:exdu:qft4} (d) has better parallelism,
which can be deduced by the awareness of different quantum gate durations. 

\section{Quantum Architecture Abstraction}
\label{sec:model}
\begin{table*}[!h]
\centering
\footnotesize
\caption{Definition of Quantum Abstract Machine.}
\label{tab:qam:def}
\begin{tabular}{l|l|l} 
\hline
&
\textbf{Notation} & \textbf{Definition} \\

\hline 
\multirow{5}{*}{\cell{Static\\ Structure}}&
\PQubit & The set of physical qubits, $|$\PQubit$|$ = $\NPQubit$;
$\forall$\pqubit{}$\in$\PQubit, \pqubit{}.\qlock is the qubit lock described in Section~\ref{sec:design:lock}\\
\cline{2-3}
&   \PGate{} & The set of elementary quantum operations and \kn{SWAP}, $|$\PGate{}$|$ = $\NPGate$\\
\cline{2-3}
&   \QCoupling=(\PQubit,\PQEdge) & The coupling 
graph of a quantum device,
\PQEdge $\subseteq$ \PQubit$\times$\PQubit \\
\cline{2-3}
&   \GDuration{}: \Arrow{\PGate{}}{\Nat} & Mapping from quantum operations to their durations, 
\Nat represents the set of natural numbers\\
\cline{2-3}
&   \QDistance: \Arrow{\PQubit$\times$\PQubit}{\Nat} & Mapping from physical qubit pairs 
to their shortest path lengths on the \QCoupling{}, \\
&& if there is no path between \pqubit{i} and \pqubit{j}, then \QDistance(\pqubit{i},\pqubit{j}) = \Intmax  \\
\hline

\multirow{2}{*}{\cell{Dynamic \\Structure}}
&   \qmap{}:\Arrow{\LQubit}{\PQubit} & Mapping from logical qubits to physical qubits \\
\cline{2-3}

&   CF(\lgates{}) & Commutative Front gate set of a gate sequence \lgates{}, defined in Definition~\ref{def:cfgates} \\
\hline

\multirow{2}{*}{\cell{Auxiliary\\Functions}}
&   \gname{\lgate{}} &  the name of a given gate \lgate{}. \\
\cline{2-3}
&   \gqubits{\lgate{}} & the logical qubit sequence applied by a given gate \lgate{}. \\
\hline

\multirow{5}{*}{\cell{Variables}}
&   \pqubitvars & Physical qubits, \pqubit{i}$\in$\PQubit, $1\le i\le \NPQubit$\\
\cline{2-3}
&   \lqubitvars & Logical qubits,  \lqubit{i}$\in$\LQubit, $1\le i\le \NLQubit$ \\
\cline{2-3}
&   \pgatevars & Physical quantum operations, \pgate{i}$\in$\PGate, $1\le i\le \NPGate$\\
\cline{2-3}
&   \lgatevars & Quantum operations in the circuit program\\
\cline{2-3}
&   \lgates{}  & A sequence of quantum operations, \lgates{} = [\lgate{1},\lgate{2},..., \lgate{k}] if $k=|$\lgates{}$|$,
and the length of \lgates{} is written as \lgates{}$.len$\\
\hline
\end{tabular}
\end{table*}
Since the qubit mapping problem is 
affected by the constraints of underlying QC devices,
which based on various and evolving quantum technologies, 
it is essential to design quantum mapping algorithms that are compatible with different quantum technologies.

\subsection{Characteristics of NISQ devices}
Table~\ref{tab:qc-args}
lists parameter information of some QC devices 
based on ion trap, superconducting and 
neutral atomic quantum technologies, respectively.
From the table,
we see that two-qubit gate executes slower 
(at least 2$\times$) 
and has lower fidelity than single-qubit gate
on both superconducting and ion trap platforms;
the ion trap system is about 1000$\times$ slower than 
the superconducting system, 
but can execute more gates before decoherence.
The directly implementable elementary gates in ion trap system are
single-qubit gate R$^\theta_\alpha$ 
(rotations by an angle $\theta$ about any axis $\alpha$) 
and two-qubit gate \kn{XX}.
Specifically, 
\kn{CNOT} gate can be implemented by
a one-\kn{XX} and four-\kn{R}~\cite{debnath2016ionqc}.
Neutral atoms are 
similar in magnitude to trapped ions.
However, 
the two-qubit gate applied to neutral atoms may not perform slower than a single-qubit gate, 
but the fidelity is much worse. \par
\subsection{Definition of the \qam}
In view of the above, 
we consider the qubit connectivity of various NISQ devices, 
and take each gate duration as a multiple of quantum clock cycle \QCycle,
which can be analogized to the classic clock cycle.
We then introduce a \qamfname 
which consists of static and dynamic structures, 
denoted as \QArch = (\QArchS, \QArchD).
Table~\ref{tab:qam:def} shows the definitions for \qam,
where \QArchS = (\PQubit, \PGate{}, \QCoupling, \GDuration{}, \QDistance),
and \QArchD = (\qmap{}, CF).
We assume the device can provide enough physical qubits (denote the number as $\NPQubit$) for the program's execution 
(denote the number of logical qubits in the program as $\NLQubit$), \ie $\NPQubit\ge\NLQubit$.

For a QC device, we abstract its qubit layout as a 
graph \QCoupling
where qubits are vertices and 
there are 
edges between 
qubit pairs where a two-qubit gate is allowed to apply on them. 
We introduce the Gate Duration Map~\GDuration{} into \QArchS
which maps each kind of quantum gate to its duration, 
depending on the information from quantum architecture. 
We assume the same kind of quantum gates have the same duration and fidelity. 
We also introduce the shortest distance matrix map \QDistance between each pair of physical qubits 
for quick selection of exchangeable qubits in our \mysys scheduling algorithm.

\par

\section{Design of the Qubit Remapping Algorithm}
\label{sec:design}
In this section, 
we discuss our design for 
\fullname.
We introduce two fundamental mechanisms that enable \mysys
context-sensitivity and duration awareness,
\ie \kw{qubit lock} 
and \kw{commutativity detection}. 
\subsection{Qubit Lock}
\label{sec:design:lock}
\mysys is based on a reasonable assumption: 
\kw{Two or more gates cannot be applied to the same qubit at the same time}. 
If a gate occupies a qubit, 
we call the qubit busy, and other gates can no longer be applied to the qubit. 
As the example shown in Fig.~\ref{fig:ex1:qft4}, 
when inserting \kn{SWAP} for a specific two-qubit gate \kn{CX q[0],q[3]}, 
the neighbour qubit \kn{q[2]} of the target qubits may be occupied by 
the contextual gate which has started in an earlier time. 
Using the occupied qubits to route the two-qubit gate will reduce the parallelism of the program because the routing process has to wait until the occupied qubits become free. 

\begin{figure}[htbp]
\centering
\begin{minipage}[b]{.25\textwidth}
    \centering
    \includegraphics[width=\textwidth]{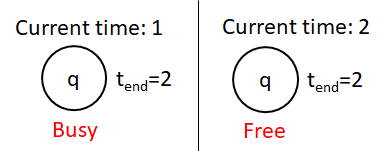}

\end{minipage}
    \caption{Qubit lock \qlock of qubit q is 2 means q is busy until time 2.
    }
    \label{fig:lock}
\end{figure}

To make \mysys aware of the qubit occupation by the past contextual gate, 
we introduce a {\bf qubit lock} \qlock for each physical qubit in \pqubit. 
When start applying a quantum gate \pgate{} $\in$ \PGate{} whose duration is \GDuration{\pgate{}} at time $t$ to a physical qubit in \pqubit{}, 
\mysys will update this qubit's \qlock as $t$+ \GDuration{\pgate{}}, which means the qubit is occupied before $t$+\GDuration{\pgate{}}. 
A qubit is \kw{free} only when its lock \qlock $\le$ current time, which means all the past gates applied on this qubit are finished. 
When trying to find a routing path for a specific two-qubit gate, 
by comparing \qlock of each qubit with the current time, 
\mysys can be aware of which qubit is occupied by the past contextual gate. 
Fig.~\ref{fig:lock} shows an example. 
Gates can only be applied to the physical qubits in a free state. 
We call the gates whose associate physical qubits are all free as \kw{lock free} gates.

Qubit lock also makes \mysys aware of the gate duration difference. 
Different gate kinds have different duration and \mysys updates the operated qubit's lock \qlock with different values. 
As a result, qubits occupied by gates with shorter duration will be set smaller \qlock 
and become free earlier. 
Thus \mysys can use those earlier free qubits to route two-qubit gates
and improve the parallelism of the program. 
As the example shown in Fig.~\ref{fig:exdu:qft4}~(d), 
suppose the program starts at time 0 and \GDuration{ T}=1, \GDuration{CNOT}=2. 
Then \qlock of \pqubit{1} is set to 1 while \qlock of \pqubit{0} and \pqubit{2} are set to 2. 
\pqubit{1} becomes free at time 1 while \pqubit{2} is still busy. 
\mysys can use \pqubit{1} to route for the third gate and need not wait for the freedom of \pqubit{2}.

\subsection{Commutativity Detection}
\label{sec:design:detect}
Qubit lock brings \mysys awareness of the past contextual gate. On the other hand,
considering gate commutation relation can expose more future contextual gate 
for \mysys to decide routing path.
\begin{definition}[Commutative Forward Gate, CF gate]
\label{def:cfgates}
    Given a gate sequence \lgates{}=[\lgate{1}, \lgate{2}, ..., \lgate{k}, ...], $\forall$\lgate{k} $\in$ \lgates{},
    \lgate{k} is a commutative forward gate iff 
    $\forall j, 0 < j < k$, \lgate{j} and \lgate{k} are commutative. 
\end{definition}
    The commutation relation between two-qubit gates \lgate{A}, \lgate{B} 
    that share qubits with each other can be resolved 
    by checking the relevant unitary operators $\hat{A}\hat{B} = \hat{B}\hat{A}$. 
    Gates applied to disjoint qubits are obviously commutative with each other.\par
    
    All the CF gates in sequence \lgates{} are denoted as \lgates{CF}.
    CF gates can be moved to the the head of \lgates{},
    which means they can be \kw{executed instantly} from logical perspective.
    Compared to the methods that 
    ignore the commutativity between quantum gates, 
    choosing CF gates as logically-executable gates can expose more contextual gates for the heuristic search 
    to determine better remapping solutions.
    
    For example, suppose a sequence \lgates{} contains two gates: \kn{CX q1,q3} and \kn{CX q2,q3} in order. 
    The second gate shares \kn{q3} with the first and might not be regarded as logically executable due to qubit dependence. 
    However, because the second commutes with the first one, both of the gates are CF gates in \lgates{} and instantly executable in fact. 
    Commutativity detection will expose both \kn{CX}s for the heuristic search 
    which improve the contextual look-ahead ability of \mysys. \par

\subsection{Algorithms for \mysys Remapping}
\label{sec:algorithm}
\begin{figure}
    \centering
    \includegraphics[width=0.48\textwidth]{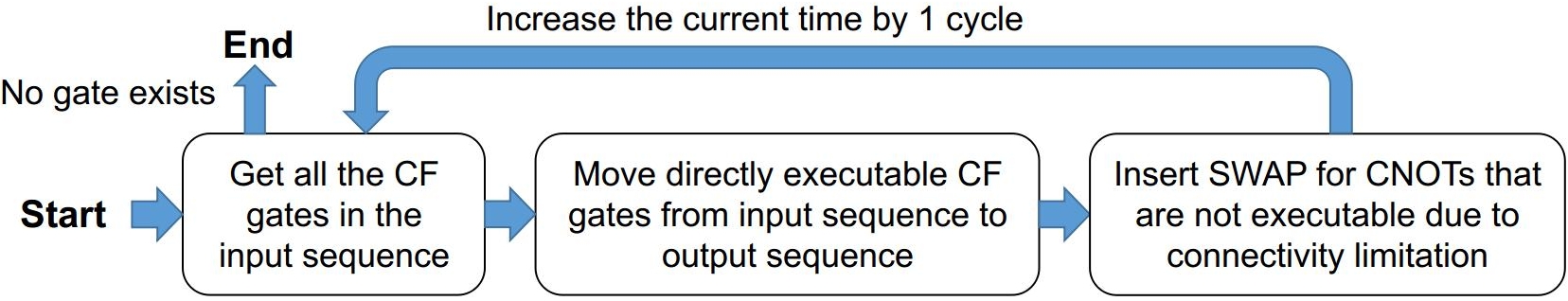}
    \caption{Overview of the remapping algorithm.}
    \label{fig:overview}
\end{figure}

Now we discuss how \mysys transforms the quantum circuit to fit the hardware limitation.
Given the coupling graph \QCoupling=(\pqubit,\PQEdge),
initial mapping \qmap{} and gate duration map \GDuration{}, 
our algorithm takes an original gate sequence \lgates{} as input, 
and generates an 
executable gate sequence \lgatesschd{}. The overview of the algorithm is shown in Fig.\ref{fig:overview}. The algorithm starts with the current time and each qubit lock initialized as 0. There are three steps in each iteration.

\mysect{Step 1}
At the start of each cycle, the algorithm first gets all CF gates in the input sequence \lgates{} denoted as \lgates{CF}. The algorithm will terminate if there is no gate in \lgates{}. 
 \mysect{Step 2}
 In this step, the algorithm will select all directly executable gates in \lgates{CF}.
 A gate \lgate{} is directly executable only when it satisfies two conditions below.
\begin{itemize}
    \item \lgate{} is a \kw{lock free} gate.
    \item If \lgate{} is a two-qubit gate like \kn{CNOT}, its two target qubits are connected in the coupling graph. 
\end{itemize}
 Then the algorithm will apply those executable gates by moving them from the input sequence \lgates{} to the output sequence \lgatesschd{} and update the qubit lock for each gate in the way described in Section~\ref{sec:design:lock}. 
\mysect{Step 3}
For the remaining \kn{CNOT}s in \lgates{CF}, 
due to the connectivity limitation, 
the algorithm need decide which \kn{SWAP}s should be inserted into the output sequence. 
The algorithm first searches all \kw{lock free} edges associated with the qubits of each two-qubit gate \lgate{} in \lgates{CF} to avoid huge overhead cost by global searching.
For example, 
suppose \lgate{} as ``\kn{CX} \lqubit{0}, \lqubit{1}'',
\pqubit{k} = \qmap{}(\lqubit{1}) is the corresponding physical qubit of \lqubit{1} and 
we locate all of \pqubit{k}'s adjacent qubits as \pqubit{k1},\pqubit{k2},...,\pqubit{kt}.
Since the number of physical qubits may be greater than that of logical qubits,
not every physical qubit has a corresponding logical qubit,
thus the \kn{SWAP} operation can only be applied to physical qubits.
If the \kn{SWAP} between \pqubit{k} and \pqubit{ki}
$(1\leq i \leq t)$ is \kw{lock free}, 
then \kn{SWAP} \pqubit{k},\pqubit{ki} will be regarded as a candidate.
We repeat this process for all qubits associated with gate \lgate{} 
to get the candidate \kn{SWAP} set denoted as \Cswap.

Next, the algorithm repeatedly select the best \kn{SWAP} in \Cswap and insert it into \lgatesschd{} until no positive-priority \kn{SWAP} remains in \Cswap. The priority of a \kn{SWAP} denotes the benefit it can bring to the circuit and we discuss how to calculate the priority for each \kn{SWAP} in next subsection. In each selection iteration, the algorithm first calculates the priority for each \pgate{swap} in \Cswap. Then the algorithm inserts the gate with the highest priority in \Cswap into \lgatesschd{} and updates the qubit locks to remove the no longer \kw{lock free} \kn{SWAP}s from \Cswap. \par

\begin{figure}
    \centering
    \includegraphics[width=0.49\textwidth]{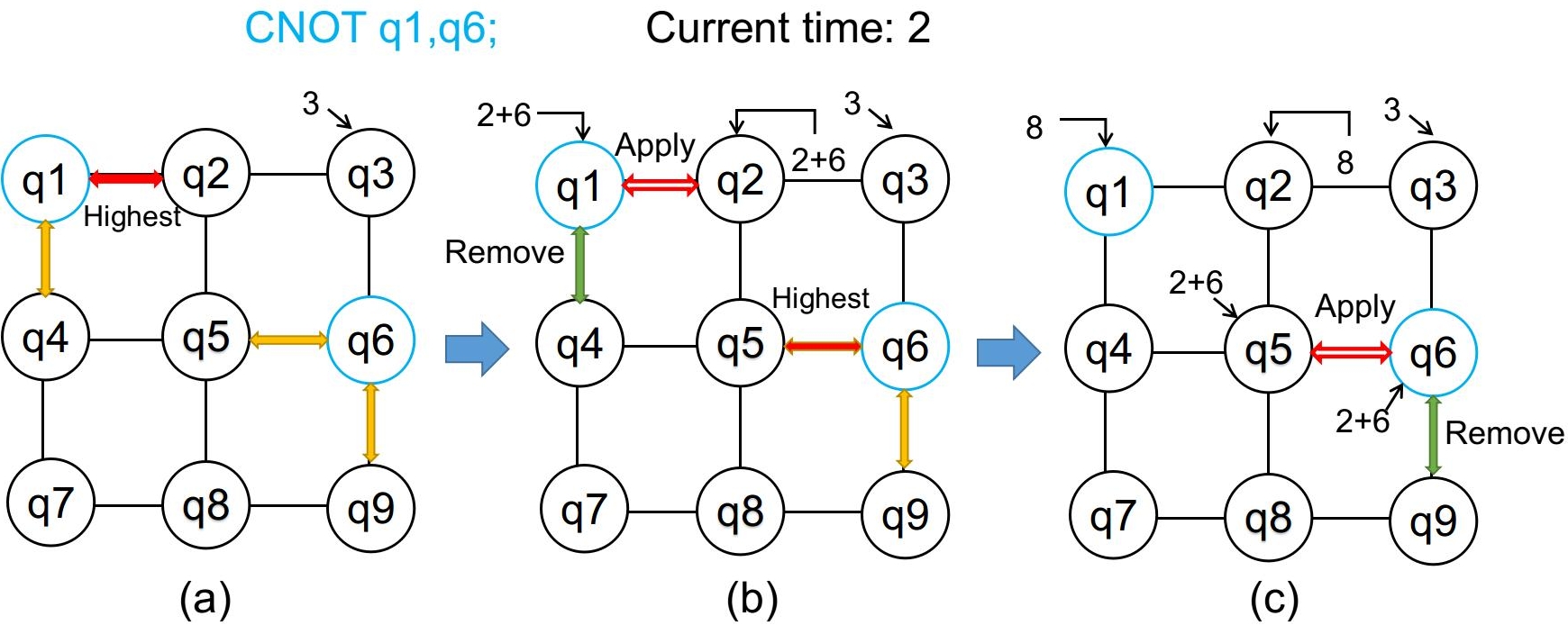}
    \caption{Example of the remapping process in one cycle. }
    \label{fig:ex-select}
\end{figure}

An example shown in Fig.\ref{fig:ex-select} illustrates the remapping process. Suppose a \kn{CNOT} between qubits \kn{q1} and \kn{q6} need be applied at cycle 2 in a 3 $\times$ 3 grid model. For an easy explanation, we suggest in this cycle all qubit locks are 0 except the lock of qubit \kn{q3} is 3, which means that q3 is busy. As shown in Fig.\ref{fig:ex-select}(a), the algorithm inspects each free edge associated with q1, q6 and calculates their priority. Because the lock of q3 is larger than the current time, the edge between q3 and q6 is not free. 
Finally, all the candidate \kn{SWAP}s are colorized and suppose the red one has the highest priority. 
Then as shown in Fig.\ref{fig:ex-select}(b), the algorithm applies the red one and updates the lock of qubit q1 and q2 (suppose the duration of \kn{SWAP} is 6 cycle). This update causes the edge between q1 and q4 is no longer free, and it is removed from \Cswap. 
The algorithm repeats the process and applies the \kn{SWAP} between q5 and q6. In the end, no gate exists in \Cswap, which leads to the termination of this cycle.\par 

\subsection{Design of Heuristic Cost Funtion}
 \label{sec:design:priority}
The heuristic cost function for SWAP \lgate{swap}
$$
Heuristic(g_{swap},\mathbb{M},\pi) = \left<
H_{\sf basic},H_{\sf fine}\right>
$$
is to measure the benefits that gate \lgate{} can bring to solve the mapping problem.
When comparing the priority between two \kn{SWAP}s, 
\Hbasic is compared first and 
\Hfine is compared only when two gates have the same \Hbasic.

\par

\mysect{Basic Priority}
Suppose a two-qubit gate \lgate{} applied to logical qubits \lqubit{1} and \lqubit{2},
\QDistance(\qmap{}(\lqubit{1}), \qmap{}(\lqubit{2})) denotes the distance of two qubits in the coupling graph \QCoupling{}. 
When the distance becomes 1, the gate fits the hardware limitation of the device. 

L(\qmap{},\lgate{})=\QDistance(\qmap{}(\lgate{}.\lqubit{1}), \qmap{}(\lgate{}.\lqubit{2})) 
calculates the distance of \lgate{} $\in$ \lgates{CF} based on the mapping \qmap{}. 
To evaluate a candidate \kn{SWAP}, 
we first temporarily use that \kn{SWAP} to update \qmap{} and get \qmap{new}. 
Then we calculate how much \qmap{new} can reduce (or increase) the distance of all \lgate{} $\in$ \lgates{CF} compared to the original \qmap{}. 
Equation~\ref{equ:basic} shows the basic priority heuristic function \Hbasic. 

\begin{equation}
    \label{equ:basic}
    H_{\sf basic} = \sum_{g\in I_{\sf CF}}L(\pi,g)-L(\pi_{new},g)
\end{equation}
If the basic heuristic function of a specific \kn{SWAP} $\le 0$,
    this \kn{SWAP} won't shorten the total distance and cannot bring any benefit to the circuit. 
    If we find no \kn{SWAP} has a positive basic heuristic, 
    there is no best \kn{SWAP} gate.
    But sometimes it happens that neither \kn{SWAP} gate nor other executable gates could be launched 
    while all qubits are free. 
    Such situations are called ``deadlock'' by us,
    and at this time we should just choose a \kn{SWAP} with the highest priority to launch, 
    even if its \Hbasic may not be positive. 

\mysect{Fine Priority}

In many cases, 
there are several candidate \kn{SWAP}s with the same basic priority, 
so we design fine priority as shown in Equation~\ref{equ:fine}
which applies to 2D lattice model. 
\begin{equation}
\label{equ:fine}
\small
    H_{\sf fine} =-\left|\VD(\pi_{new}, g)-\HD(\pi_{new}, g)\right|
\end{equation}
Functions \HD and \VD stand for the horizontal and vertical distance 
on the lattice between the two qubits of \lgate{}.
The main idea of fine priority is: 
suppose a two-qubit gate \lgate{} with specific distance $D = \HD + \VD$, 
there may be $\mathrm{C}_{D}^{\HD} = \frac{\HD!\VD!}{(\HD+\VD)!}$,
possibly the shortest routing way for \lgate{}, 
which increases as $\left|\VD-\HD\right|$ decreases. 
Choose a suitable \kn{SWAP} gate to make $\left|\VD-\HD\right|$ smaller 
may let the remapping algorithm perform better in the future.

\begin{figure}
    \centering
    \includegraphics[width=0.23\textwidth]{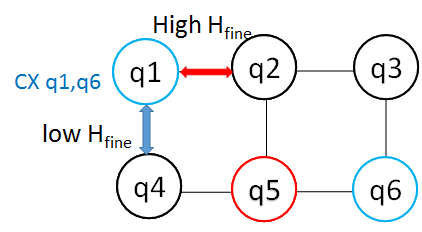}
    \caption{The circuit is going to apply \kn{CX} on q1 and q6 while q5 will be locked for a long time. The \kn{SWAP} between \{q1,q2\} and between \{q1,q4\} have the same \Hbasic. But routing q1 through q4 will be blocked by q5 while routing through q2 can get better parallelism.}
    \label{fig:ex:fine}
\end{figure}

Retaining more possible routing way can improve the parallelism. 
Fig.\ref{fig:ex:fine} shows a situation:
 \Hfine indicates the algorithm to choose \kn{SWAP} 
 that can let the \kn{CX} gate has closer vertical and horizontal distance 
 which can avoid waiting for \lqubit{5} to be free.

\subsection{Example}
\label{sec:design:ex}
\begin{figure}
    \centering
\includegraphics[width=0.49\textwidth]{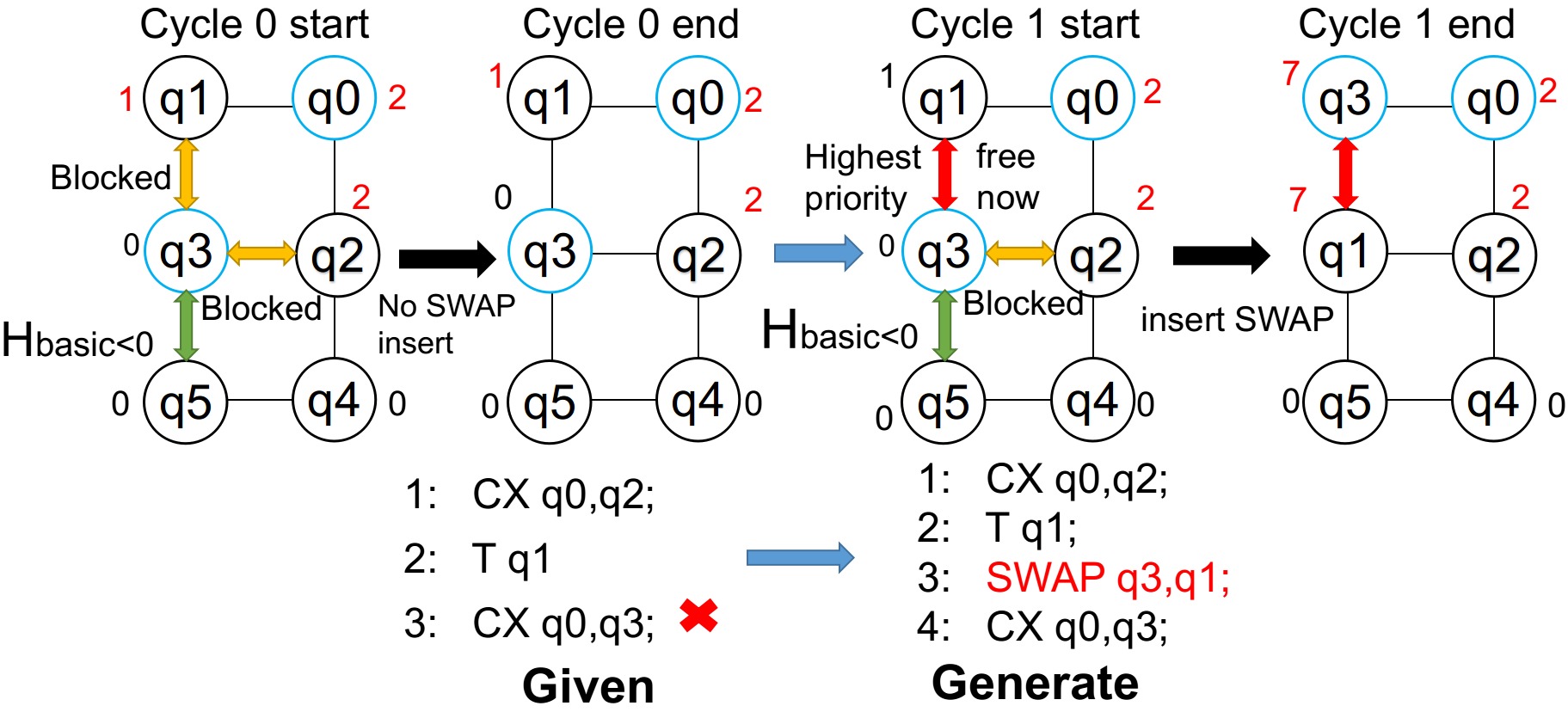}
\caption{Example of heuristic search for the high parallelism \kn{SWAP}. The number near the qubits denotes the qubit lock \qlock.}
\label{fig:ex-alg}
\end{figure}

Now we use an example shown in Fig.~\ref{fig:ex-alg} to explain our algorithm. 
Suppose there is a 6-qubit device and we are given a gate sequence \lgates{} 
that contains a \kn{CX} on \{q0,q2\}, a \kn{T} on \{q1\} and a \kn{CX} on \{q0,q3\}. 
The number near the qubit node represents the value of its \qlock. 
All the three gates are CF gates. 
Due to the coupling limitation, \kn{CX} on \{q0,q3\} is not directly executable and 
\kn{SWAP} is needed. 
The algorithm simulates the execution timeline and starts at cycle 0. 

At cycle 0, the first gate ``\kn{CX} q0,q2'' and the second gate ``\kn{T} q1'' are directly executable 
so both of them will be launched and qubits \{q0,q1,q2\}'s \qlock locks are updated with the gate duration
(\kn{T}=1 cycle, \kn{CX}=2 cycle). 
Each of \{q0,q1,q2\} has bigger \qlock than current time 
and thus they are locked. 
Therefore the \kn{SWAP} between \{q1,q3\} and \{q2,q3\} are blocked. 
\kn{SWAP} between \{q3,q5\} with \Hbasic$<0$
(which means the \kn{SWAP} won’t shorten the total distance of CF gates according to our heuristic cost function)
moves q3 away from q0 and 
will not be inserted. 
As a result, no \kn{SWAP} will be inserted in cycle 0 and 
the mapping \qmap{} stays unchanged. 

At cycle 1, qubit q1 becomes free while q2 stays busy. 
The \kn{SWAP} between \{q1,q3\} becomes free while the \kn{SWAP} between \{q3,q2\} is still blocked. 
Therefore the algorithm knows that the \kn{SWAP} between \{q1,q3\} can start earlier than \kn{SWAP} between \{q3,q2\} 
and choose \kn{SWAP} q3,q1 to solve the remapping problem. 
After launching ``\kn{SWAP q1,q3}'', 
qubit locks of \{q1,q3\} are also updated by the sum of its start time (cycle 1) and 
the duration of \kn{SWAP} (6 cycle) as 7.

\section{Experimental Evaluation}
\label{sec:eval}
In this section, 
we evaluate \mysys with benchmarks based on the latest reported hardware models. 
\mysect{Comparison with Previous Algorithms}
Several recent algorithms proposed by IBM~\cite{qiskit},
Siraichi $\etal$\cite{siraichi2018qubitalloc}, 
Zulehner $\etal$\cite{zulehner2019mapping} and Li $\etal$\cite{li2019tackling} try to find solutions of the qubit mapping problem with small circuit depth. 
Among them, 
Li's SABRE~\cite{li2019tackling} beats the other three in the performance of benchmarks,
thus it is used for comparison in this paper. \mysect{Hardware Configuration}
We test our algorithm on several latest reported architectures, including IBM Q20 Tokyo\cite{li2019tackling}, IBM Q16 Melbourne\cite{ibmqdeviceinfo}, $6\times6$ grid model
proposed by Enfield~\cite{siraichi2018qubitalloc}'s GitHub and Google Q54 Sycamore~\cite{arute2019google54}. 
The gate duration difference configuration is based on experimental data of symmetric superconducting technology 
shown in Table~\ref{tab:qc-args}, 
where two-qubit gate duration is generally twice as much as that of the single-qubit gate. 
\mysect{Benchmarks}
To evaluate our algorithm, we totally collect 71 benchmarks which are selected from the previous work, including:
1) programs from IBM Qiskit~\cite{cross2018ibmqiskit}'s Github and  RevLib~\cite{wille2008revlib};
2) several quantum algorithms compiled from ScaffCC~\cite{abhari2014scaffcc} and Quipper~\cite{green2013quipper};
3) benchmarks used in the best-known algorithm SABRE~\cite{li2019tackling}. 
The size of the benchmarks ranges from using 3 qubits up to using 36 qubits and about 30,000 gates.
For the IBM Q16, Q20 and $6\times6$ architectures,
68 benchmarks out of the 71 benchmarks except 3 36-qubit programs are tested. 
While all 71 benchmarks are tested on Google Q54 Sycamore. 
%
%

\newcommand{\figwidth}{5.6cm}
\newcommand{\figheight}{4cm}
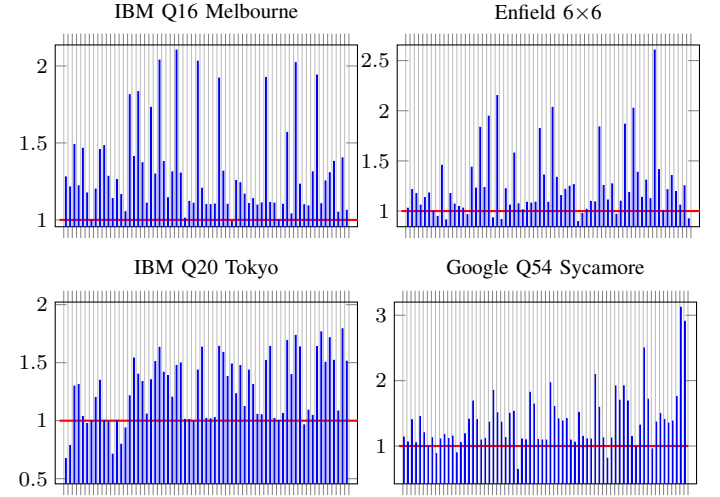
\begin{figure}[h]
\footnotesize
\begin{minipage}[b]{0.24\textwidth}
\centering
\begin{tikzpicture}
\begin{axis}[
ybar,
title=IBM Q16 Melbourne,
width=\figwidth,
height=\figheight,
enlargelimits=0.03,
xticklabels=\empty,
ybar interval=0.2,
]
\draw[red,thick] (axis cs:0,1) -- (axis cs:71,1);
\addplot coordinates {
(1,1.279)(2,1.214)(3,1.490)(4,1.221)(5,1.465)
(6,1.175)(7,1.000)(8,1.200)(9,1.457)(10,1.482)
(11,1.283)(12,1.138)(13,1.262)(14,1.164)(15,1.053)
(16,1.813)(17,1.411)(18,1.833)(19,1.371)(20,1.109)
(21,1.731)(22,1.298)(23,2.038)(24,1.379)(25,1.144)
(26,1.312)(27,2.103)(28,1.304)(29,1.011)(30,1.120)
(31,1.108)(32,2.031)(33,1.205)(34,1.099)(35,1.099)
(36,1.103)(37,1.921)(38,1.316)(39,1.101)(40,0.989)
(41,1.256)(42,1.241)(43,1.167)(44,1.106)(45,1.138)
(46,1.096)(47,1.110)(48,1.925)(49,1.113)(50,1.109)
(51,1.000)(52,1.101)(53,1.568)(54,1.039)(55,2.021)
(56,1.232)(57,1.098)(58,1.091)(59,1.312)(60,1.941)
(61,1.105)(62,1.253)(63,1.306)(64,1.380)(65,1.050)
(66,1.403)(67,1.062)(68,1.000)
};

\end{axis}
\end{tikzpicture}
\end{minipage}
~
\begin{minipage}[b]{0.24\textwidth}
\centering
\begin{tikzpicture}
\begin{axis}[
ybar,
title=Enfield 6$\times$6,
width=\figwidth,
height=\figheight,
enlargelimits=0.03,
xticklabels=\empty,
ybar interval=0.2,
]
\draw[red,thick] (axis cs:0,1) -- (axis cs:71,1);
\addplot coordinates {
(1,1.028)(2,1.214)(3,1.174)(4,1.059)(5,1.134)
(6,1.180)(7,1.001)(8,0.947)(9,1.456)(10,0.912)
(11,1.174)(12,1.069)(13,1.045)(14,1.029)(15,0.964)
(16,1.437)(17,1.229)(18,1.835)(19,1.234)(20,1.946)
(21,0.932)(22,2.151)(23,0.915)(24,1.222)(25,1.059)
(26,1.578)(27,1.072)(28,1.012)(29,1.087)(30,1.079)
(31,1.089)(32,1.822)(33,1.359)(34,1.087)(35,2.033)
(36,1.336)(37,1.154)(38,1.217)(39,1.247)(40,1.263)
(41,0.895)(42,0.976)(43,1.017)(44,1.096)(45,1.091)
(46,1.839)(47,1.255)(48,1.110)(49,1.271)(50,0.967)
(51,1.098)(52,1.866)(53,1.184)(54,2.025)(55,1.386)
(56,1.136)(57,1.308)(58,1.123)(59,2.605)(60,1.413)
(61,1.004)(62,1.213)(63,1.354)(64,1.194)(65,1.059)
(66,1.252)(67,0.922)(68,1.102)
};
\end{axis}
\end{tikzpicture}
\end{minipage}

\begin{minipage}[b]{0.24\textwidth}
\centering
\begin{tikzpicture}
\begin{axis}[
ybar,
title=IBM Q20 Tokyo,
width=\figwidth,
height=\figheight,
enlargelimits=0.03,
xticklabels=\empty,
ybar interval=0.2,
ymin=0.5,
]
\draw[red,thick] (axis cs:0,1) -- (axis cs:71,1);
\addplot coordinates {
(1,0.673)(2,0.786)(3,1.298)(4,1.312)(5,1.036)
(6,0.975)(7,1.000)(8,1.200)(9,1.348)(10,1.000)
(11,0.989)(12,0.711)(13,1.000)(14,0.796)(15,0.937)
(16,1.214)(17,1.540)(18,1.400)(19,1.337)(20,1.057)
(21,1.353)(22,1.509)(23,1.632)(24,1.417)(25,1.389)
(26,1.202)(27,1.476)(28,1.497)(29,1.011)(30,1.011)
(31,0.997)(32,1.435)(33,1.633)(34,1.019)(35,1.016)
(36,1.026)(37,1.640)(38,1.588)(39,1.380)(40,1.488)
(41,1.232)(42,1.475)(43,1.122)(44,1.435)(45,1.312)
(46,1.054)(47,1.051)(48,1.519)(49,1.640)(50,1.019)
(51,1.000)(52,1.062)(53,1.690)(54,1.397)(55,1.735)
(56,1.636)(57,0.964)(58,1.090)(59,1.045)(60,1.639)
(61,1.765)(62,1.503)(63,1.715)(64,1.520)(65,1.083)
(66,1.792)(67,1.512)(68,1.978)
};
\end{axis}
\end{tikzpicture}
\end{minipage}
~~~~
\begin{minipage}[b]{0.24\textwidth}
\centering
\begin{tikzpicture}
\begin{axis}[
ybar,
title=Google Q54 Sycamore,
width=\figwidth,
height=\figheight,
enlargelimits=0.03,
xticklabels=\empty,
ybar interval=0.2,
ymin=0.5,
]
\draw[red,thick] (axis cs:0,1) -- (axis cs:71,1);
\addplot coordinates {
(1,1.139)(2,1.063)(3,1.404)(4,1.051)(5,1.451)
(6,1.202)(7,1.000)(8,1.125)(9,0.885)(10,1.107)
(11,1.174)(12,1.114)(13,1.152)(14,0.897)(15,1.053)
(16,1.188)(17,1.411)(18,1.688)(19,1.404)(20,1.091)
(21,1.115)(22,1.363)(23,1.850)(24,1.507)(25,1.363)
(26,1.130)(27,1.500)(28,1.530)(29,0.643)(30,1.108)
(31,1.096)(32,1.821)(33,1.641)(34,1.099)(35,1.091)
(36,1.091)(37,1.967)(38,1.604)(39,1.414)(40,1.382)
(41,1.420)(42,1.092)(43,1.060)(44,1.515)(45,1.148)
(46,1.110)(47,1.105)(48,2.094)(49,1.592)(50,1.125)
(51,0.813)(52,1.121)(53,1.919)(54,1.701)(55,1.921)
(56,1.689)(57,1.145)(58,0.988)(59,1.320)(60,2.500)
(61,1.715)(62,0.956)(63,1.366)(64,1.498)(65,1.406)
(66,1.349)(67,1.382)(68,1.756)(69,3.123)(70,2.908)
(71,2.081)
};
\addplot coordinates{};
\end{axis}
\end{tikzpicture}
\end{minipage}

    \caption{Speedup ratio of all 71 benchmarks compared between \mysys and SABRE in four architectures. The benchmarks are listed from left to right in the ascending order of the number of qubits used.}
    \label{fig:result}
\end{figure}
\subsection{Circuit Execution Speedup}
We collect the weighted circuit depth of the circuits produced by \mysys and SABRE for the 71 benchmarks. Initial mapping has been proved to be significant for the qubit mapping problem, and for a fair comparison, we use the same method as SABRE to create the initial mapping for the benchmarks. 
We use the depth of circuits produced by SABRE compared with the one of \mysys to show the ability of our algorithm to speed up the quantum program. 
As shown in Fig.\ref{fig:result}, 
the average speedup ratio of \mysys on four architecture models, IBM Q16 Melbourne, Enfield 6$\times$6, IBM Q20 Tokyo and Google Q54 are respectively 1.212, 1.241, 1.214 and 1.258.

\subsection{Fidelity Maintenance}
\begin{figure}[h]
\begin{minipage}[b]{0.24\textwidth}
    \centering
    \includegraphics[width=\textwidth]{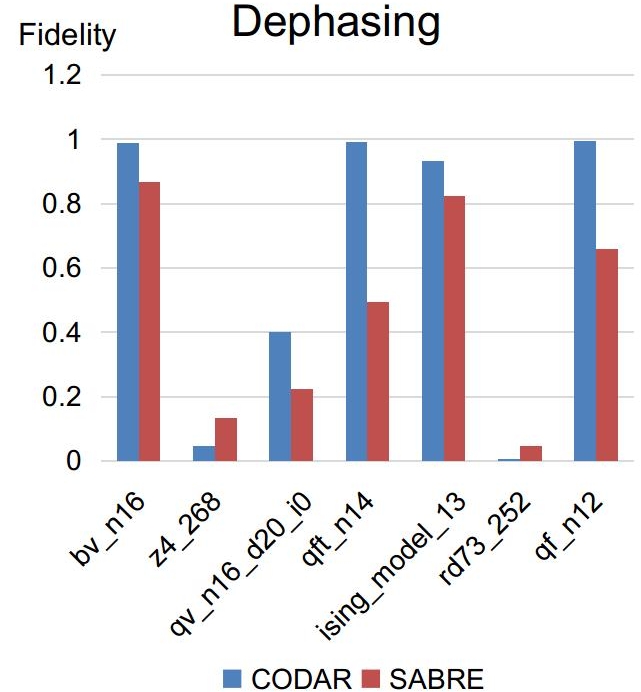}
\end{minipage}
~
\begin{minipage}[b]{0.24\textwidth}
    \includegraphics[width=\textwidth]{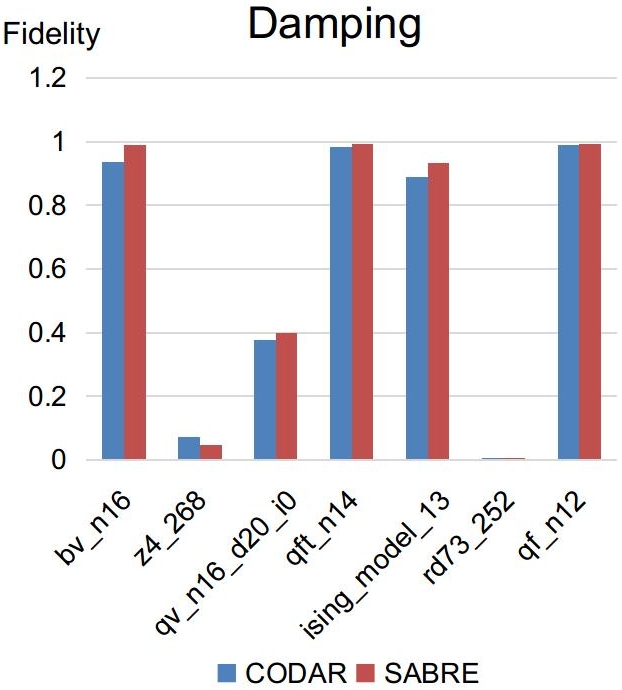}
\end{minipage}
    \caption{Fidelity of the circuit produced by \mysys and SABRE for several quantum algorithms. When the noise mainly caused by qubit dephasing, \mysys performs better than SABRE and the fidelity of several circuits produced by \mysys reach nearly 1. When the noise mainly caused by qubit damping, \mysys performs about the same with SABRE.  }
    \label{fig:fidelity}
\end{figure}

Fidelity has been proved to be significant for the quantum computer in NISQ era. \mysys focuses on exploring the parallelism of the program to reduce the execution time and ignores the number of \kn{SWAP}s inserted into the program. Compared to SABRE, \mysys may insert more \kn{SWAP}s, which may bring more noise to the program. However, less execution time will improve the fidelity of the circuit on the contrary. To show \mysys's ability to maintain the fidelity, we use a distributed noisy quantum virtual machine made by Origin Quantum~\cite{qpanda},
which is based on Qubit Dephasing and Damping model~\cite{nielsen2010quantum:qcqi} to simulate the fidelity of 7 famous quantum algorithms. 
The result (Fig.\ref{fig:fidelity}) indicates that \mysys can speed up the circuits and maintain the fidelity of the circuits at the same time. 
\section{Conclusion}
\label{sec:concl}
In NISQ era, most quantum programs are not directly executable 
because two-qubit gates can be applied between arbitrary two logical qubits 
while it can only be implemented between two adjacent physical qubits due to hardware constraints. 
To solve this problem,
In this paper we propose \mysys that can transform the original circuit and insert necessary \kn{SWAP} operations 
making the circuit comply with the hardware constraints. 
With the design of qubit lock and commutativity detection, 
\mysys is aware of the program context and the gate duration difference, 
which help \name find the remapping with good parallelism and
reduce QC's weighted depth. 
Experimental results show that 
\name can speed up quantum programs by 1.212$\sim$1.258 in different quantum architectures
compared with the best known algorithm
on average and maintain the fidelity of the benchmarks when running on OriginQ quantum noisy simulator. 

\section*{Acknowledgement}
{\footnotesize
This work was partly supported by the grants of Anhui Initiative in Quantum Information Technologies (No. AHY150100), the National Natural Science Foundation of China (No. 61772487) and Anhui Provincial Major Teaching and Research Project (No. 2017jyxm0005). The authors would like to thank Prof. Peng Xu \etal from Wuhan Institute of Physics and Mathematics for their enthusiastic answers to the consultation on neutral atomic quantum technologies;
and also thank Origin Quantum Computing Technology Co., Ltd. (OriginQ) for providing quantum noisy simulation platform for testing.
}

\begin{scriptsize}
\iffalse
\input{main.bbl}
\else
\bibliography{quantum}
\bibliographystyle{plain}
\fi
\end{scriptsize}
\end{document}